\documentclass[preprint,prd,aps,nofootinbib]{revtex4}
\usepackage{amsmath}
\usepackage{psfig}
\newcommand{\be}{\begin{equation}}
\newcommand{\ee}{\end{equation}}
\newcommand{\ba}{\begin{eqnarray}}
\newcommand{\ea}{\end{eqnarray}}
\newcommand{\bc}{\begin{center}}
\newcommand{\ec}{\end{center}}

\def\cstok#1{\leavevmode\thinspace\hbox{\vrule\vtop{\vbox{\hrule\kern1pt
\hbox{\vphantom{\tt/}\thinspace{\tt#1}\thinspace}}
\kern1pt\hrule}\vrule}\thinspace}
\begin{document}
\title{Gribov Problem for Gauge Theories: a Pedagogical
Introduction}
\author{Giampiero Esposito$^{1,2}$
\thanks{Electronic address: giampiero.esposito@na.infn.it}
Diego N. Pelliccia$^{3,4}$
\thanks{Electronic address: diego.pelliccia@fe.infn.it}
Francesco Zaccaria$^{2,1}$
\thanks{Electronic address: zaccaria@na.infn.it}}
\affiliation{
${ }^{1}$Istituto Nazionale di
Fisica Nucleare, Sezione di Napoli,\\
Complesso Universitario di Monte S. Angelo, Via Cintia,
Edificio N', 80126 Napoli, Italy\\
${ }^{2}$Dipartimento di Scienze Fisiche,
Complesso Universitario di Monte S. Angelo,\\
Via Cintia, Edificio N', 80126 Napoli, Italy\\
${ }^{3}$Universit\`a di Ferrara,
Dipartimento di Fisica,\\
Via Paradiso 12, 44100 Ferrara, Italy\\
${ }^{4}$Istituto Nazionale di
Fisica Nucleare, Sezione di Ferrara,\\
Via Paradiso 12, 44100 Ferrara, Italy}

\vspace{2cm}
\begin{abstract}
The functional-integral quantization of non-Abelian gauge theories
is affected by the Gribov problem at non-perturbative level: the
requirement of preserving the supplementary conditions under gauge
transformations leads to a non-linear differential equation, and
the various solutions of such a non-linear equation represent
different gauge configurations known as Gribov copies. Their
occurrence (lack of global cross-sections from the point of view
of differential geometry) is called Gribov ambiguity, and is here
presented within the framework of a global approach to quantum
field theory. We first give a simple (standard) example for the
$SU(2)$ group and spherically symmetric potentials, then we
discuss this phenomenon in general relativity, and recent developments,
including lattice calculations.
\end{abstract}
\maketitle
\section{Introduction}
In the modern geometrical formulation of field theories one may start
with general properties of classical dynamical systems described
globally by a trajectory or history [1]. A {\it history} is a section
of a fibre bundle $E$ having the space-time manifold $(M,g)$ as its
base space. The typical fibre is known as {\it configuration space}
and can be denoted by $C$. If the dynamical system consists of a set of
fields, such a set takes its values in $C$. One usually
assumes that a fibre bundle $E$ is associated with a group ${\cal G}$;
such a group can act globally on the fields, or
instead it can be a Lie group that acts locally on each fibre. In the
latter case, it is more appropriate to consider a principal
fibre bundle $P$ [2] which has ${\cal G}$ itself as the typical fibre, and a
connection on $P$. The connection on $P$ defines a connection 1-form
on the space-time $(M,g)$, known as a {\it gauge field} and taking
values in the Lie algebra $L({\cal G})$ of
${\cal G}$. The connection is therefore a
collection of sections of trivial bundles $U_{i} \times T^{*}(M)
\times L({\cal G})$, which are related to each other by group actions in
the overlaps $U_{i} \cap U_{j}$. These group actions are called
{\it gauge transformations}, and for the whole dynamics to remain
invariant under gauge transformations the gauge field itself has to
become dynamical. The space $\Phi$ of all possible field histories
includes both those that do and those that do not satisfy the
Euler--Lagrange equations (when there is an action principle
for the former), and can be viewed as an
infinite-dimensional manifold. To sum up, the space of histories
$\Phi$ is a principal fibre bundle having the infinite-dimensional
Lie group ${\cal G}$, the gauge group, as its typical fibre, while
physics takes place on the base space of this bundle, the latter
being the quotient space $\Phi / {\cal G}$ called the space of 
orbits [1].

On the other hand,
the functional-integral quantization of gauge theories of fundamental
interactions relies upon the following properties.
For a given gauge theory with action functional $S$ and generators
of infinitesimal gauge transformations $R_{\; \alpha}^{i}$ with associated
linearly independent vector
fields ${\cal R}_{\alpha} \equiv R_{\; \alpha}^{i} \;
{\delta \over \delta \varphi^{i}}$, one has (hereafter, Greek indices
from the beginning of the alphabet are Lie-algebra indices, Greek indices 
from the middle of the alphabet are space-time indices, while lower case
Latin indices are used for fields at a space-time point [1])
\begin{equation}
{\cal R}_{\alpha}S={\cal R}_{\beta}S=0 \Longrightarrow
\Bigr[{\cal R}_{\alpha},{\cal R}_{\beta}\Bigr]S
={\cal R}_{\alpha}{\cal R}_{\beta}S
-{\cal R}_{\beta}{\cal R}_{\alpha}S=0 .
\label{(1)}
\end{equation}
If all flows that leave the value of the action invariant can be
expressed, at each point of $\Phi$, as linear combinations of the 
${\cal R}_{\alpha}$ and skew fields at that point, the most general
solution of Eq. (1) reads as
\begin{equation}
\Bigr[{\cal R}_{\alpha},{\cal R}_{\beta}\Bigr]
=C_{\; \alpha \beta}^{\gamma} \; {\cal R}_{\gamma}
+{\delta S \over \delta \varphi^{j}} T_{\; \alpha \beta}^{j}.
\label{(2)}
\end{equation}
Hereafter, we assume that the tensor fields $T_{\alpha \beta}$
vanish, and that the
$C_{\; \alpha \beta}^{\gamma}$ are constant, in that they
are independent of the fields (but they might depend on
space-time coordinates). The $C_{\; \alpha \beta}^{\gamma}$ are
therefore structure constants of an infinite-dimensional Lie
group. Since we are studying a gauge theory,
what is crucial is the existence of equivalence classes under
the action of gauge transformations. The information on such
equivalence classes is encoded into suitable `coordinates',
say $(I^{A},\chi^{\alpha})$, obtained as follows. The $I^{A}$
are {\it non-local functionals} which pick out the orbit where
the field $\varphi$ lies, whereas the $\chi^{\alpha}(\varphi)$
pick out the particular point on the orbit corresponding to the
field $\varphi$. They are precisely the gauge functionals
considered by Faddeev and Popov [3]. The physical
$\langle {\rm out} | {\rm in} \rangle$ amplitude can be
expressed, formally, by functional integration over equivalence
classes (under gauge transformations) of field configurations:
\begin{equation}
\langle {\rm out} | {\rm in} \rangle = \int \mu(I)
{\rm e}^{{\rm i}S(I)}{\rm d}I .
\label{(3)}
\end{equation}
It is now possible, at least formally, to re-express this
abstract functional-integral formula in terms of the original
field variables, under the assumption (not always verified) that
the $\chi^{\alpha}$ coordinates are globally defined.
For this purpose, the integration over
equivalence classes is made explicit by introducing the
$\chi$-integration with the help of a $\delta$ distribution,
say $\delta(\chi(\varphi)-\zeta)$. After obtaining the Jacobian
$J(\varphi)$ of the coordinate transformation from
$(I^{A},\chi^{\alpha})$ to $\varphi^{l}$, its functional
logarithmic derivative shows that $J$ takes the form
$J(\varphi)=N(\varphi){\rm det}{\cal F}$, where a careful use of
dimensional regularization can be applied to reduce
$N(\varphi)$ to a factor depending only on the non-local
functionals $I^{A}$ [1]. Its effect is then absorbed
into the measure over the field configurations, so that the
$\langle {\rm out} | {\rm in} \rangle$ amplitude reads as
\begin{equation}
\langle {\rm out} | {\rm in} \rangle = \int {\rm d}\varphi \;
{\rm det}\,{\cal F} \; \delta(\chi(\varphi)-\zeta) {\rm e}^{{\rm
i}S[\varphi]} . \label{(4)}
\end{equation}
At this stage one performs a Gaussian average over all gauge
functionals, and denoting by $\rho_{\alpha \beta}$ a constant
invertible matrix, and by $\sigma^{\alpha}$ 
and $\psi^{\beta}$ two real-valued and
{\it independent} fermionic fields [1], the functional integral
for the $\langle {\rm out} | {\rm in} \rangle$ amplitude is
eventually re-expressed in the form
\begin{equation}
\langle {\rm out} | {\rm in} \rangle
= \int {\rm d}\varphi \; {\rm d}\sigma \; {\rm d}\psi \;
{\rm e}^{{\rm i}[S(\varphi)
+\chi^{\alpha}\rho_{\alpha \beta}\chi^{\beta}
+\sigma_{\alpha}{\cal F}_{\; \beta}^{\alpha}\psi^{\beta}]} .
\label{(5)}
\end{equation}
The physical predictions of the theory are $\chi$- and
$\rho$-independent.
The operator ${\cal F}_{\; \beta}^{\alpha}$ is called the ghost operator,
and the fields $\sigma^{\alpha}$ and 
$\psi^{\beta}$ are the corresponding ghost
fields [1]. The physical idea is due to Feynman [4], while the
development of the corresponding formalism in quantum gravity was
obtained by DeWitt [5]. However, to have a well defined
functional-integral representation, it is more convenient to start
from an Euclidean formulation. In particular, in the one-loop
quantum theory, one considers infinitesimal gauge transformations,
and if the theory with action $S$ is bosonic, one tries to choose
$\chi^{\alpha}$ in such a way that both the ghost operator and the
operator on perturbations of the gauge field are of Laplace type
[6]. The addition of the gauge-averaging (or gauge-fixing) term
$\chi^{\alpha}\rho_{\alpha \beta}\chi^{\beta}$ plays a crucial
role in ensuring that both the ghost operator and the gauge-field
operator have a well defined Green function. Such a term
corresponds to the supplementary (or gauge-fixing) condition
already occurring in the classical theory [5].

\section{Asymptotic conditions for pure Yang--Mills theories}

In geometrical language, the Yang--Mills potentials 
$A_{\; \mu}^{\alpha}(x)$ are components of the pull-back
to space-time of the Lie-algebra-valued 
connection 1-form $A_{\; \mu}^{\alpha}{\rm d}x^{\mu}$
on the principal Yang--Mills bundle, and one
has basis matrices $G_{\alpha}$ for an $l$-dimensional 
irreducible representation
of the Yang--Mills Lie algebra with structure constants $f_{\; \beta
\gamma}^{\alpha}$ such that
$$
[G_{\alpha},G_{\beta}]=G_{\gamma} \; f_{\; \alpha \beta}^{\gamma}.
$$
The gauge potentials
$A_{\; \mu}^{\alpha}$ and basis matrices $G_{\alpha}$ 
can be used to define the matrices
$A_{\mu} \equiv G_{\alpha} \; A_{\; \mu}^{\alpha}$,
and one can write finite gauge
transformations in the form
\begin{equation}
{ }^{D}A_{\mu} \equiv -D_{,\mu}D^{-1}
+DA_{\mu}D^{-1}.
\label{(6)}
\end{equation}
This equation describes the gauge orbit for the configuration
$A_{\mu}$, i.e. the set of all gauge-equivalent configurations
corresponding to $A_{\mu}$.
One can further say that finite {\it little gauge transformations}
[1] are the proper subset of transformations described
by Eq. (6) such that the matrix $D$ can be written as
\begin{equation}
D={\rm e}^{G_{\alpha}\xi^{\alpha}},
\label{(7)}
\end{equation}
where the $\xi^{\alpha}$ are finite functions on space-time that
vanish at spatial infinity. This condition implies that $D$ tends to
the identity therein. The matrices $D$ are the representation
matrices for the group $\cal G$ generated by the $G_{\alpha}$.

{\it Big gauge transformations} are defined by Eqs. (6) and (7),
supplemented by the following asymptotic conditions on the
$D$ matrices (the $\xi^{\alpha}$ functions no longer vanish
at spatial infinity):
\begin{equation}
D(x) \rightarrow \pm 1_{l} \; {\rm as} \; x \rightarrow \infty,
\label{(8)}
\end{equation}
$1_{l}$ being the $l$-dimensional unit matrix.
In this case, both the identity and minus the identity are necessary,
because the transformation law (6) does not change on replacing $D$
by $-D$.

More precisely, the infinite-dimensional Lie group ${\cal G}$ with
structure constants $C_{\; \alpha \beta}^{\gamma}$
as in Eq. (2), with representation
matrices expressed by Eq. (7), is the {\it proper gauge group}
(its elements are the little gauge transformations defined above). The
{\it full gauge group} is obtained by adjoining to the proper gauge
group all transformations of the space $\Phi$ of histories into itself,
independent of field variables, which leave the action functional $S$
un-affected and do not result from global symmetries.  
When big gauge transformations exist, both the full gauge group and
the space of histories have disconnected components.

Note that both big and little gauge transformations do not affect the
dynamical variables at infinity. This leads to an investigation of
the boundary conditions that should be imposed upon the dynamical
variables. In general, the Yang--Mills fibre bundle might be non-trivial
in spacelike directions. The possible physical implications of such
a non-trivial nature have not yet been studied, and we shall assume
hereafter that the space-time manifold is asymptotically Minkowskian,
and that non-trivial aspects can only occur in the time direction [1].

On focusing on a four-dimensional space-time, one requires that the
components of the Yang--Mills potential should have the following
asymptotic behaviour [1]:
\begin{equation}
A_{\; 0}^{\alpha} \sim {1\over r^{1/2 +\varepsilon}}, \; \; \; \; \; \;
A_{\; i}^{\alpha} \sim {1\over r^{1+\varepsilon}},
\label{(9)}
\end{equation}
where $\varepsilon$ is an arbitrarily small positive number. Moreover,
the time derivatives of $A_{i}^{\alpha}$ are taken to
satisfy the fall-off condition
\begin{equation}
A_{\; i,0}^{\alpha} \sim {1\over r^{3/2 +\varepsilon}},
\label{(10)}
\end{equation}
which implies that the field strength defined as [1]
\begin{equation}
F_{\; \mu \nu}^{\alpha} \equiv A_{\; \nu , \mu}^{\alpha}
-A_{\; \mu , \nu}^{\alpha}+f_{\; \beta \gamma}^{\alpha} \;
A_{\; \mu}^{\beta} \; A_{\; \nu}^{\gamma}
\label{(11)}
\end{equation}
satisfies the fall-off conditions
\begin{equation}
F_{\; i0}^{\alpha} \sim {1\over r^{3/2 +\varepsilon}}, \; \; \; \; \; \;
F_{\; ij}^{\alpha} \sim {1\over r^{2+\varepsilon}}.
\label{(12)}
\end{equation}
Three properties are therefore found to hold [1]:
\vskip 0.3cm
\noindent
(i) All physically admissible solutions of the field equations are
included in the space $\Phi$ of field histories, in particular
those having non-vanishing total charge. For these solutions,
the charge integral converges.
\vskip 0.3cm
\noindent
(ii) The spatial integral of the energy density $T^{00}$ is finite,
whether or not the field strength $F_{\mu \nu}^{\alpha}$ in the
expression of the energy-momentum tensor $T^{\mu \nu}$ satisfies the
field equations.
\vskip 0.3cm
\noindent
(iii) The spatial integral of the Yang--Mills Lagrangian is finite.

The above conditions are compatible with Eq. (8), but add a more
accurate requirement: one should require the conditions
\begin{equation}
D \mp 1_{l} \sim {1\over r^{\varepsilon}}, \; \; \; \; \; \;
D_{,0} \sim {1\over r^{1/2 +\varepsilon}}.
\label{(13)}
\end{equation}
The space $\Phi$ consists therefore of all 
bundle structures of field histories
$A_{\mu}^{\alpha}(x)$ satisfying the asymptotic conditions (9) and
(10) in every patch of the fibre bundle. The gauge group $\cal G$,
which is the typical fibre of $\Phi$, can be identified with the set
of all matrix functions, having the exponential form in Eq. (7) and
satisfying the asymptotic conditions (8) and (13). The transition
functions between patches of the bundle are matrix functions of
the same form, restricted to overlapping regions. Last, note that
the vanishing trace of the generators $G_{\alpha}$ implies that
${\rm det} \; D(x)=1$ for all $x$.

In the Hamiltonian formalism, a frequently used supplementary
condition is the {\it axial gauge}. The resulting ghost operator
reads as [1]
\begin{equation}
{\cal F}_{\; \beta'}^{\alpha}=
-\delta_{\; \beta}^{\alpha} \delta_{,1}(x,x')
-f_{\; \gamma \beta}^{\alpha}A_{\; 1}^{\gamma}\delta(x,x'),
\label{(14)}
\end{equation}
the Green function of which, with indices suppressed for
simplicity, is
\begin{eqnarray}
G(x,y)&=& \Bigr[a \theta(x^{1},y^{1})+(a-1)\theta(y^{1},x^{1})\Bigr]
\delta(x^{0},y^{0}) \delta(x^{2},y^{2}) \delta(x^{3},y^{3})
\nonumber \\
& \times & {\cal P} \; {\rm exp}\left[-\int_{y^{1}}^{x^{1}}
f_{\alpha}A_{1}^{\alpha}(x^{0},z^{1},x^{2},x^{3}){\rm d}z^{1}
\right],
\label{(15)}
\end{eqnarray}
where $a$ is any real number, and $\cal P$ denotes path ordering [1].
Moreover, the $f_{\alpha}$ are matrices built out of the structure
constants $f_{\; \alpha \gamma}^{\beta}$, such that
$$
[f_{\alpha},f_{\beta}]=f_{\gamma} \; f_{\; \alpha \beta}^{\gamma},
$$
and they generate the adjoint representation
of the gauge group [1].

One of the apparently attractive features of these Green functions
is their local nature in the time variable, i.e. ghost fields do not
propagate. However, the axial gauge is incompatible with the asymptotic
conditions (9), (10), (12), (13) which are necessary 
to define the space of field
histories and the gauge group with a certain control of the limiting
behaviour at spatial infinity.

To prove the latter statement, let us consider the standard
gauge-fixing condition $\chi^{\alpha}=0$. One can assume that, to fulfill
such a condition, one has only to perform a finite gauge transformation
(see Eq. (6)) such that
$$
0={ }^{D}A_{1}=-D_{,1} \; D^{-1}+D \; A_{1} \; D^{-1},
$$
or, equivalently, $D_{,1}=D \; A_{1}$. This equation can be solved on
each patch of the fibre bundle, by simply integrating with the
condition that $D=1_{l}$ at $x^{1}=-\infty$. In general, however, $D$ does
not wind up equalling the identity at $x^{1}=\infty$ for all
$x^{0},x^{2}$ and $x^{3}$, even if the potentials
$A_{\; \mu}^{\alpha}$ satisfy the asymptotic conditions (9). The same
occurs when the $\chi^{\alpha}$ are set equal to any other functions
of the $x^{\mu}$ variables that vanish at spatial infinity [1].

For a supplementary (or gauge) condition to be compatible with the
asymptotic conditions (9), (10), (12), (13) 
the resulting Green functions of ghost
fields should be non-local in the time variable. For example, the
temporal gauge $A_{\; 0}^{\alpha}=0$ is instead compatible with the
asymptotic conditions given above [1].

\section{Gribov phenomenon}

The careful assignment of asymptotic conditions on gauge potentials
is justified by the analysis of global properties of the latter.
In the case of Yang--Mills theories, a new phenomenon is found to
occur with respect to quantum electrodynamics.

Since the discovery of instantons [7], the interest in global properties
led to the attempt of understanding quark confinement as well by
virtue of such global properties [8,9], while perturbative methods failed to
provide any clue on such aspects of gauge theories. Among the peculiar
features of global nature, the {\it Gribov ambiguity} deserves special
attention. In the simplest possible terms, on requiring that the
supplementary (or gauge-fixing) condition should be preserved under
gauge transformations of the potential, one obtains a non-linear
differential equation, {\it for each choice of gauge-fixing compatible
with the asymptotic conditions} (9), (10) and (13). The key remark is now
that, on the space of orbits ${\cal O} \equiv 
{\Phi / {\cal G}}$ which is the base
space of $\Phi$, the choice of gauge-fixing on the potentials:
$$
\chi[A]-\zeta=0
$$
describes a certain surface
which intersects [10] the orbits of Eq. (6).  
In the very definition of gauge fixing at
perturbative level, one requires that each orbit should intersect the
gauge-fixing surface only once, so as to ensure uniqueness of the
potentials satisfying dynamical equations for a given choice of gauge
fixing. In the attempt of extending such a procedure globally all over
${\cal O}$, however, one faces difficulties resulting from the non-trivial
topology of the fibre bundle in the non-Abelian case [11]. Thus, the
gauge-fixing surface intersects some orbits more than once. In other
words, in the overlapping regions among various patches of the bundle,
transition functions are not single-valued; they might remain single-valued,
however, within each topological sector. {\it If} this is the case, the
perturbative evaluation of in-out amplitudes through the gauge-fixed
functional integral is justified. 
By contrast, at global level, transition functions
may take different values in the various topological sectors, by virtue
of particular gauge transformations which relate them. By virtue of
these {\it big gauge transformations}, two potentials close to each
other in a given patch with respect to the metric on the space
of histories [1]
$$
\gamma_{\alpha \beta'}^{\; \mu \nu'} \equiv K_{YM}^{-2}
\sqrt{-g} \; \gamma_{\alpha \beta} \; g^{\mu \nu} \delta(x,x'),
$$
where $K_{YM}$ is the Yang--Mills coupling constant,
$\gamma_{\alpha \beta} \equiv -{\rm tr}(f_{\alpha}f_{\beta})$ 
is the Cartan--Killing metric and $g^{\mu \nu}$ is the contravariant
form of the space-time metric tensor,
can turn out to be quite far apart in another patch, since one of the two can
now lie in a different topological sector.

Indeed, the Gribov phenomenon can be viewed as a topological
obstruction to achieving continuity in the choice of values
$\zeta^{\alpha}$ taken by the gauge-fixing functional
$\chi^{\alpha}$. In geometrical language, one can say that it is
impossible to find a global cross-section for the Yang--Mills
fibre bundle. Singer [12] has indeed proved that, for $SU(N)$
groups with base manifold $S^{3}$ or $S^{4}$, no gauge-fixing
condition exists which is compatible with the asymptotic
conditions (9), (10), (12), (13). 
Since the more involved symmetry groups
admit $SU(2)$ as a sub-group, the above statement holds for all
physically relevant gauge theories. It should be pointed out that,
in the case of quantum chromodynamics, two particular gauge-fixing
conditions exist for which the Gribov ambiguity does not seem to
occur: the axial and temporal gauges. However, the former does not
satisfy the asymptotic conditions 
(9), (10), (12), (13) as we have seen at the
end of Sec. II, while in the latter
the Gauss law does not automatically hold. If the Gauss law is
also enforced, one again finds the occurrence of Gribov copies
[13].

Along the lines of the original Gribov argument [14],
let us consider a non-Abelian
theory with potentials subject to the infinitesimal gauge
transformations
\begin{equation}
\delta A_{\; \mu}^{\alpha} \equiv \delta_{\beta}^{\alpha}
\partial_{\mu}\lambda^{\beta}
+f_{\; \gamma \beta}^{\alpha} \; A_{\; \mu}^{\gamma} \; \lambda^{\beta}
\equiv D_{\; \mu \beta}^{\alpha} \; \lambda^{\beta}.
\label{(16)}
\end{equation}
Moreover, let us impose a covariant gauge condition such as the Lorenz
gauge for Yang--Mills theory:
\begin{equation}
\partial^{\mu}A_{\; \mu}^{\alpha}=0,
\label{(17)}
\end{equation}
for potentials satisfying the asymptotic conditions (9) and (10).
The general field variables $\varphi^{i}$ of Sec. I are now the Yang--Mills
potentials $A_{\; \mu}^{\alpha}(x)$, and Eq. (17) is an example of
gauge choice implemented, in the path integral (4), through the
Dirac distribution $\delta(\chi(\varphi)-\zeta)$.
The requirement of preserving the gauge-fixing condition (17)
under gauge transformations yields
\begin{equation}
0=\partial^{\mu}\Bigr(A_{\; \mu}^{\alpha}+\delta A_{\; \mu}^{\alpha}\Bigr)
-\partial^{\mu}A_{\; \mu}^{\alpha}
=\partial^{\mu}D_{\; \mu \beta}^{\alpha} \; \lambda^{\beta}
={\cal F}_{\; \beta}^{\alpha} \; \lambda^{\beta}.
\label{(18)}
\end{equation}
This can be seen as an equation defining the singularities of the
ghost operator, i.e. its zero-modes. More precisely, one can interpret
Eq. (18) as an eigenvalue equation for the ghost operator with
vanishing eigenvalue. If the corresponding eigenfunction is non-trivial,
it is said to be a {\it zero-mode}, and the ghost operator becomes singular.
A physics-oriented interpretation is instead as follows: there exist
a number of gauge potentials, related to each other by a gauge
transformation, and satisfying the dynamical equations with a given
choice of supplementary (or gauge-fixing) condition, when the ghost fields
$\psi^{\beta}$ have zero-mass bound states. The latter interpretation is
suggested by the fact that ${\cal F}_{\; \beta}^{\alpha}$ occurs in the
action functional of the theory only through the term
${{\rm i} \sigma_{\alpha} {\cal F}_{\; \beta}^{\alpha} 
\psi^{\beta}}$ in Eq. (5).
Equation (18) is therefore equivalent to having
\begin{equation}
{\cal F}_{\; \beta}^{\alpha}[A_{\mu}]\psi^{\beta}
=\varepsilon \; \psi^{\alpha}, \; {\rm with} \;
\varepsilon=0,
\label{(19)}
\end{equation}
where the parameter $\lambda^{\beta}$ has been replaced by the
ghost field $\psi^{\beta}$.

On assuming, for simplicity, that the gauge potential vanishes before
performing the gauge transformation, Eq. (19) becomes, for a generic
value $\varepsilon$ of the ghost eigenvalue,
\begin{equation}
\partial^{\mu}\partial_{\mu} \; \delta_{\; \beta}^{\alpha}
\; \psi^{\beta}=\delta_{\; \beta}^{\alpha} \; \cstok{\ } \psi^{\beta}
=\varepsilon \; \psi^{\alpha}.
\label{(20)}
\end{equation}
This equation can only be solved for non-negative values of
$\varepsilon$. On taking increasingly large magnitudes of the
potentials, i.e. on moving away from the neighbourhood of
$A_{\mu}=0$ where perturbative calculations are performed for
Yang--Mills theories, one arrives at a sufficiently large
magnitude of $A_{\mu}$ for which there exist solutions of Eq. (19)
with $\varepsilon =0$, i.e. Eq. (18) is satisfied. On further
increasing the magnitude of the potentials one finds negative
eigenvalues, until for an even bigger value of such a magnitude
one finds again a solution of Eq. (19) with $\varepsilon =0$.

Note that Eq. (20) is, at this stage, 
an hyperbolic equation having non-vanishing
solutions also in the limiting Abelian case; a way to get rid of
non-uniqueness of the solutions of the equation expressing preservation
of the gauge-fixing is to perform a Wick rotation and hence consider
an Euclidean space. In such a space one can understand the behaviour
of the solutions of Eq. (20) by exploiting index theory and the
concept of spectral flow (see Ref. [15] and references therein).

One can thus imagine that the potentials solving Eq. (18) divide the
space of potentials into regions, in each of which Eq. (20) has
a certain number of eigenvalues, corresponding to bound states of the
ghost field $\psi^{\beta}$. In correspondence to the surfaces defined
by the potentials  satisfying Eq. (18), known as {\it Gribov horizons},
which separate the various {\it Gribov regions}, there exist massless
ghost states. The first Gribov region $C_{0}$ has no ghost bound
states, the second Gribov region $C_{1}$ has one ghost bound state,
the $k$-th Gribov region $C_{k}$ has $(k-1)$ ghost bound states.

Since the potentials occurring in the Gribov regions $C_{n}$,
$n=1,2,3{\ldots}$ are Gribov copies of the configurations occurring
in $C_{0}$, as Gribov himself did show in his original paper, the
functional integration should be restricted to $C_{0}$. Two kinds of
Gribov copies can be found, i.e. those obtained from the potentials
in $C_{0}$ from {\it big gauge transformations}, resulting from
topological effects previously described, as well as equivalent copies
of the potentials in $C_{0}$ that exist within $C_{0}$ itself.
While the former can be simply ignored in perturbative calculations,
the latter make it necessary to use a greater care. To avoid
overcounting copies in the functional integral, it is necessary to
further restrict integration to a domain known as the
{\it fundamental modular region} [16], the boundary of which is studied
in Refs. [16-18], and within which the Gribov ambiguity no 
longer occurs [19].

In the original analysis by Gribov [14], the restriction of the
integration region in configuration space has the effect of removing
the infrared singularity of perturbative theory and leads to a linear
increase of the interaction among non-Abelian charges at large
distances, i.e. a possible mechanism of quark confinement.

\section{Gribov equation}

Before further discussing the implications of the Gribov analysis,
it is appropriate to present a simple example, used by Gribov himself,
of the occurrence of such a phenomenon. An explicit solution of
the equation expressing preservation of the Coulomb
gauge-fixing condition at $A_{\mu}=0$ under
gauge transformations:
\begin{equation}
\partial_{r}((\partial^{r}D)D^{-1})=0,
\label{(21)}
\end{equation}
can be found for the $SU(2)$ group, which can be
easily parametrized because the exponential map on its Lie algebra is
in one-to-one correspondence with the interior of the smallest circle
where all points map to $-I$. Thus,
every element with the exception of minus the identity has a
unique representation $U$:
\begin{equation}
U={\rm e}^{{\rm i}{\vec n} \cdot {\vec \sigma}}=
I \cos |{\vec n}|+{\rm i}{\hat n} \cdot {\vec \sigma}
\sin |{\vec n}|,
\label{(22)}
\end{equation}
where ${\hat n} \equiv {\vec n} / |{\vec n}|$ and
${\vec n} \cdot {\vec n}< \pi$. To obtain an element $U({\vec x})$ of
the gauge group, one has simply to make $\vec n$ into a function of
$\vec x$. Let us now assume, to obtain radial symmetry and the correct
asymptotic behaviour, that [20]
\begin{equation}
{\vec n}={\omega(r)\over r}(y,x,z),
\label{(23)}
\end{equation}
where the exchange of $x$ and $y$ is intended, and let us pass
to spherical polar coordinates, obtaining
\begin{equation}
U(r,\theta,\phi)=
\begin{pmatrix}
c_{\omega}+{\rm i}c_{\theta}s_{\omega}
\hfill & {\rm e}^{{\rm i}\phi} s_{\theta} s_{\omega} \hfill \\
-{\rm e}^{-{\rm i}\phi} s_{\theta} s_{\omega} \hfill &
c_{\omega}-{\rm i}c_{\theta}s_{\omega} \hfill
\end{pmatrix}
\label{(24)}
\end{equation}
where $s_{\alpha}$ and $c_{\alpha}$ denote $\sin \alpha$ and
$\cos \alpha$, respectively. To obtain Eq. (24) one has to consider
the Jacobian matrix for the coordinate transformation and the
standard row $\times$ column rule for the scalar product
${\hat n} \cdot {\vec \sigma}$. On using the familar expressions
of the divergence and gradient operators in polar coordinates,
and the further change of coordinate $t \equiv \log(r)$, Eq. (21) on
$U$ yields [20]
\begin{equation}
\left(U_{tt}+U_{t}+U_{\theta \theta}+{c_{\theta}\over s_{\theta}}
U_{\theta}+{1\over s_{\theta}^{2}}U_{\phi \phi}\right)
U^{\dagger}+U_{t}U_{t}^{\dagger}
+U_{\theta}U_{\theta}^{\dagger}+{1\over s_{\theta}^{2}}
U_{\phi}U_{\phi}^{\dagger}=0,
\label{(25)}
\end{equation}
where subscripts denote derivative with respect to the variable
expressed. In the light of Eq. (24), Eq. (25) takes the form
\begin{equation}
\begin{pmatrix}
{\rm i} c_{\theta} \hfill & {\rm e}^{{\rm i}\phi}s_{\theta}
\hfill \\
-{\rm e}^{-{\rm i}\phi}s_{\theta} \hfill &
-{\rm i}c_{\theta} \hfill
\end{pmatrix}
\Bigr({\ddot \omega}+{\dot \omega}-2 \sin \omega \cos \omega \Bigr)=0,
\label{(26)}
\end{equation}
where a dot denotes derivative with respect to $t$. On passing to
the variable ${\overline \omega} \equiv 2 \omega$, and leaving aside
the $2 \times 2$ matrix in Eq. (26) (since it is independent of
$\overline \omega$), one finds eventually a differential equation
describing a damped pendulum in a constant gravitational field, with
$\overline \omega$ being the angle from the point of unstable
equilibrium. Since $U=I$ at the origin, one finds
${\overline \omega}=2k \pi$, where $k$ is an integer. Moreover,
since $\overline \omega$ is determined up to a multiple of $2 \pi$,
there is no loss of generality in taking $k=0$. One therefore obtains
what is called the {\it Gribov pendulum equation}
\begin{equation}
\left({{\rm d}^{2}\over {\rm d}t^{2}}+{{\rm d}\over {\rm d}t}
-2 \sin \right){\overline \omega}=0,
\label{(27)}
\end{equation}
with the following boundary (or asymptotic) condition:
\begin{equation}
\lim_{t \to -\infty}{\overline \omega}(t)=0.
\label{(28)}
\end{equation}
There exist three solutions with the initial condition (28):
either the pendulum never falls, or it falls anti-clockwise, or instead
clockwise. 

An analogous treatment of the Gribov pendulum equation, which however
does not rely upon the matrix representation, can be found in Ref. [21].
At a deeper level, {\it more than a solution is found because the whole
of configuration space is being explored}. On restricting the analysis
to a bounded region, as it always occurs in perturbation theory, Gribov
copies are instead ignored.

\section{Gribov ambiguity and general relativity}

In Sec. II, on discussing the asymptotic
conditions for gauge theories, we focused our attention on
Yang--Mills theories. Actually, the manifestly covariant formalism can
treat gravitation as well in the same language used for non-Abelian
gauge theories. It may be interesting to investigate a possible
occurrence of the analogue of Gribov ambiguity in general
relativity. Several ways to express Einstein's theory as a gauge
theory have been proposed in the literature; 
the matter is that one can treat this theory
just in the connection formalism adopted in the introduction for any
dynamical system.

General relativity can be mathematically formulated starting from a
pseudo-Riemannian 4-manifold ${\cal M}$ endowed with an atlas of
coordinate charts. Symmetries usually considered are
transformations of the metric tensor $g_{\mu\nu}$ induced by
general coordinate transformations, hence the passive point of view
of the diffeomorphism group is preferred to the active one. The
biggest group $Q$ of passive dynamical symmetries of Einstein's
equations is the group of transformations reading as [22]
$$
{x'}^{\mu}=f^{\mu}(x^{\nu},g_{\rho \sigma}(x)).
$$
It is larger than the
general coordinate transformation group, usually considered, which
is a non-normal subgroup of $Q$ [23].

In a gauge theory of
gravitation one can adopt a principal fibre bundle which has,
as base space, the
union of equivalence classes of all metric tensors, solutions of
Einstein's equations, generated by passive diffeomorphisms [23].
Despite previous considerations about a more general symmetry
group involved, such a choice of principal
bundle is sufficient to characterize the
theory. The quotient group where dynamics is reduced, in fact, is
the same for the three symmetry groups: the passive
diffeomorphisms, the active ones or $Q$. The elements of this
quotient group are the gauge orbits of general relativity. It is
natural to investigate topological properties in this case, seeking
for the existence of global cross-sections, whose lack is found
in Yang--Mills theory.

The differences from the latter theory result from the 
nature of the diffeomorphism
group, which is a non-analytic infinite-dimensional Lie group,
while Yang--Mills symmetry groups are Baker--Campbell--Hausdorff
groups [24]. The Gribov ambiguity arises 
from the non-trivial topology of
gauge orbits in Yang--Mills theory; in general relativity
their topology is even less regular, so that the problem becomes 
more involved as well as more fundamental. 

Mathematical complications arise from the nature of the diffeomorphism group.
Yang--Mills theory is a field theory on a background space-time,
and the action of the gauge group in that case is on a inner space.
In general relativity, instead, the 
action of the gauge group is an extension to tensors 
over space-time of the diffeomorphisms of space-time itself [25].

It is very difficult, in the manifestly covariant configuration
space approach, to face the Gribov problem in general relativity. Indeed, 
one would need tools to make a clean separation between
gauge variables and a basis of gauge-invariant observables. In
other words, it is not simple to find suitable coordinates
$(I^{A}, \chi^{\alpha})$ as in Sec. I. 
The Hamiltonian formulation has, at least
locally, a natural tool for such a separation, i.e. the Shanmugadhasan
canonical transformations [26]. In this approach, the singularity of
Lagrangians, both in particle physics and in general relativity, 
makes it necessary to use 
the Dirac--Bergmann theory [27] of constrained Hamiltonian 
systems. Since it is
the analogue of the space of gauge orbits in the global approach,
only the constraint submanifold of phase space has physical relevance.

In phase space there are as many arbitrary Hamiltonian gauge
variables as first-class constraints, which are the 
generators of the Hamiltonian
gauge transformations. They may be used as the $I^{A}$ coordinates
of Sec. I, hence they determine a
coordinatization of the gauge orbits inside the constraint
submanifold. To obtain the reduced phase space, one has to 
add as many gauge-fixing constraints as first-class ones.
As a consequence of imposing gauge-fixing constraints,
first-class constraints are turned into the second class
and therefore treated with the help of Dirac brackets
(such constraints are then strongly vanishing). 

The Dirac observables, which in general can be non-local,
only give a coordinatization of the physically relevant space. 
On the other hand, it is the analysis of its topological properties 
that makes it possible to say 
whether a given dynamical system (with
constraints) admits a subfamily of globally defined Shanmugadhasan canonical
transformations. The existence of such a subfamily is the same 
as the one of a global cross-section, from the point of view 
of differential geometry: 
it means that the system admits preferred global separations 
between gauge and observable
degrees of freedom. In other words, in the Hamiltonian approach the
existence of globally defined Shanmugadhasan canonical
transformations makes it possible to avoid the Gribov ambiguity.

Actually, the existence of these transformations has not yet been
proved. Indeed, important constraints like the
Yang--Mills Gauss law and the ADM supermomentum constraints [25]
are partial differential equations of ellyptic type, hence they may
admit zero modes according to the choice of 
function space. One has to solve the same kind of problems
previously treated about the ghost operator in the global approach. 

To obtain, at least as a
first approximation, the main non-topological properties of a
system, the Hamiltonian approach needs to avoid the analogue of
Gribov ambiguity in general relativity. In fact, in this formalism
one assumes that all fields have to belong to suitably
weighted Sobolev spaces so that the allowed spacelike
hypersurfaces are Riemannian 3-manifolds without Killing vectors.

We may conclude that in general relativity one has to face 
the analogue of the Gribov problem; in this theory it is more
difficult to arrive at some conclusion by virtue of highly non-trivial
topology of the diffeomorphism group.

\section{Recent developments in continuum gauge theories}

We have seen in Sec. II that, to control the asymptotic
behaviour of gauge potentials, one has to impose appropriate
boundary conditions. The choice of these conditions rules the
occurrence of Gribov copies. However, one could make a distinction
between two different cases of ambiguity of gauge potentials,
according to which kind of boundary conditions are imposed. In
this way one can have a weak Gribov problem and a strong one [28].

When there exist always some configurations which have copies,
we can talk of the {\it weak Gribov problem}. 
This phenomenon, which is found in any regular gauge,
occurs even on a compact space and inside a given 
topological sector of configuration space.

When there are copies of the vanishing configuration 
$(A_{\mu}=0)$, hence pure
gauge potentials satisfying the gauge condition, we 
can talk of the {\it strong Gribov problem}. 
On choosing weak-decay boundary conditions, we might find 
the occurrence of this phenomenon in an infinite space.
On the other hand, it is not customary inside a 
given topological sector in a compact space;
the same occurs on choosing strong decay conditions in a non-compact space. 

In terms of functional integration the weak problem is less
harmful than the strong one. In fact, one might view the strong
Gribov ambiguity as a lack of strict positivity for 
the functional measure, and the weak
one as a lack of monotonicity. For smooth configurations on
a smooth compact four-dimensional space, the strong Gribov 
problem is not expected.

We have seen that boundary conditions are fundamental in 
dealing with the Gribov ambiguity. 
Indeed, appropriate boundary conditions (or compactification) 
may be imposed to avoid the strong Gribov problem, as
is usually done in the constructive study of the ultraviolet limit of
non-Abelian gauge theories to avoid the infrared
problem. The weak Gribov phenomenon, instead, cannot be avoided [28].
Anyway, in perturbative gauge theories, a topological analysis shows
that the Gribov problem is irrelevant [29].

One could manage to get rid of the Gribov ambiguity by employing different
theories, that generalize the usual gauge theories. For generalized
connections [30], the Gribov problem is completely irrelevant for
the calculation of functional integrals, hence for the  
$\langle {\rm out} | {\rm in} \rangle$ amplitude, on assuming 
absolute continuity of the
considered measure on the space of Ashtekar's generalized
connections [31] with respect to the induced Haar measure.

Up to this limitation, the Gribov phenomenon is still present
with generalized connections, but it is not a problem.
Indeed, we have found that topological non-triviality of 
the configuration space
is responsible for the occurrence of Gribov ambiguity; this non-triviality
is concentrated on a zero-measure subset [30] in this formalism.

Among several techniques proposed to take care of Gribov copies, one 
might use a quantization which does not require gauge fixing [32]
such as a generalized Gupta--Bleuler method [33], one can modify
either the gauge-fixing procedure [34] or the functional integration
[35], otherwise using non-local variables such that gauge fixing is
no longer required [36].

Another recently proposed way to get rid of the Gribov ambiguity is a
restriction on the norm of the $\langle {\rm out} |
{\rm in} \rangle$ amplitude, not considering its phase. In this
way Yang--Mills theory would be free from ambiguities if used for
the description of observables [37].

The most promising way to avoid the Gribov ambiguity in QCD is,
as far as we can see, the
stochastic quantization approach [38], in which, although there
are Gribov copies, they have no influence on expectation values.
This approach determines an Euclidean probability distribution
directly in configuration space, i.e. the space of gauge
potentials, without reduction to the orbit space. 

The first Gribov
region may be characterized as the set of relative minima with
respect to local gauge transformations of the minimizing
functional $F_{A}[D] \equiv ||{ }^{D}A||^{2}$, analogue of the ghost
operator ${\cal F}$ of Sec. III. The fundamental modular region $\Lambda$,
instead, may be characterized as the set of absolute minima. The
latter is free of Gribov copies, apart from the identification of
gauge-equivalent points on the boundary $\partial\Lambda$, and may
be identified with the gauge orbit space.

One can clearly say that in continuum gauge theories the Gribov
ambiguity is still an open problem at non-perturbative level.
In this respect, one should here mention an important result of
Neuberger [39], according to which the action used in BRST quantization
may engender vanishing partition functions, by virtue of Gribov copies
which cancel each other's effects, and the same may occur for
expectation values of physical operators.

Apart from future developments of the previous methods or other ones to
get rid of the Gribov phenomenon, for the time being some conclusions may
be outlined about implications of the existence of Gribov copies
in continuum gauge theories.

First, in the high-energy sector of non-Abelian gauge theory, 
the restricted domain of integration in the functional
amplitude has surprising consequences: they contradict the expected standard
behaviour of the theory [28]. Moreover, the Gribov problem 
leads to a cutoff on frequencies stronger than the
expected one, resulting from standard perturbative asymptotic freedom
[28]. 

In the infrared sector, Gribov himself in his original paper argued
that this ambiguity leads to an effective infrared cutoff; 
in fact, he related this
behaviour with confinement. Last, we may conclude by saying
that in non-Abelian theory the Gribov phenomenon seems to be an obstruction,
at non-perturbative level, to the coexistence of too many energy
scales [28].

\section{The case of lattice gauge theories}

In the formulation of lattice gauge theories [40], the gauge-fixing
procedure and the generation of Gribov copies, obtained via
Monte Carlo simulations and, in particular, via random gauge
rotations [41], are by now well-established techniques, although they
have been initially ignored since it is not strictly necessary to
fix a supplementary (or gauge) condition in such theories. The
gauge-fixing procedure, however, is particularly convenient in some
circumstances, and these have led to an investigation of the Gribov
problem in this framework as well [42].

The regularization provided by a lattice [43], in fact, makes
the gauge group compact, so that the Gibbs average of any
gauge-invariant quantity is well-defined. However,
asymptotic freedom causes the continuum limit to be the weak coupling
limit, and such an expansion requires gauge fixing.
Moreover, to extract non-perturbative results from Monte Carlo
simulations, one can use the evaluation of quark/gluon matrix elements,
which again requires gauge fixing.
The same procedure is also used in smearing techniques. 

A brief discussion of the lattice theory formalism may be useful
to understand why Gribov copies also occur in lattice gauge
theory. One can expect their presence by virtue of the Killingback
analysis [44], which is analogous to the investigation of Singer,
but is made for Euclidean gauge theories with periodic boundary
conditions.

An action describing fermions and gauge fields on the
lattice is obtained by replacing differentials with finite differences
and constructing a suitable covariant derivative. The theory
without fermions is known as ``pure gauge theory''.
The pure gauge action on the lattice is usually taken to be [45]
\begin{equation}
S = \beta \sum_{p} \left(
1-\frac{1}{N_{C}}\, {\rm Re}\, {\rm Tr}\,U_{p}\,\right), 
\label{(29)}
\end{equation}
where $N_{C} = 2$ for $SU(2)$, and the {\it plaquette variable}
$U_{p}$ is the ordered product of four link variables in a square.
The link variables $U_{ij}$ are the degrees of freedom in the
theory, and are elements of the gauge group. 
A link between two nearest-neighbour sites $i$ and
$j$ of the lattice is associated with each link 
variable $U_{ij}$. Such a variable
is also often written as
$U_{\mu}(x)$, where $x$ denotes the site of the link and $\mu$ the
direction.

Elements of the gauge group are the gauge transformation variables too,
which live on the lattice sites themselves. A gauge
transformation acts on a link variable as follows:
\begin{equation}
U^{\prime}_{ij} = D_{i}\,U_{ij}\,D_{j}^{-1}.
\label{(30)}
\end{equation}

On using the fundamental representation for gauge fields, the
action (29) is known as the {\it Wilson action}. The first term in
the expression is a constant and is added in order to reduce the action
to the Yang--Mills one in the classical continuum
limit. This limit is taken by letting the lattice spacing go to
zero after identifying
\begin{equation}
U_{ij} = {\rm exp} \left[ -(x^{j}-x^{i})_{\mu}\,A^{\mu}\,\left(
\frac{x^{j}+x^{i}}{2}\right)\,\right]. 
\label{(31)}
\end{equation}

Gauge fixing on the lattice requires to minimize the functional
\begin{equation}
F_{L}[D] = -\sum_{\mu, x} {\rm Re}\,{\rm Tr}
\,D(x)\,U_{\mu}(x)\,D^{\dagger}(x+\hat{\mu}),
\label{(32)}
\end{equation}
as a function of all gauge transformations $D(x)$. The
sum over $\mu$ runs from 1 to 3 for the Coulomb gauge, and from 1
to 4 for the Landau gauge. Note that this function is the lattice
analogue of the functional $F_{A}[D]=||{ }^{D}A||^{2}$. 

On expanding $U_{\mu}(x)$ according to the formal identity [46]
\begin{equation}
U_{\mu}(x) = {\rm exp}(-a\,A_{\mu}(x))=1 -a\,A_{\mu}(x) +
a^{2}A_{\mu}^{2}(x) + \dots, 
\label{(33)}
\end{equation}
where $a$ is the lattice spacing, one gets
\begin{equation}
F_{L}[1] = -\sum_{\mu, x} {\rm Tr}\,(1 + a^{2}A_{\mu}^{2}(x))=
{\rm const.} + a^{2}F_{A}[1], 
\label{(34)}
\end{equation}
where use has been made of the identity Tr$\,A_{\mu} = 0$. 
Hence, it is clear that
minimizing $F_{L}[D]$ is the same as minimizing $F_{A}[D]$,
in the classical continuum limit.

The functional $F_{L}[D]$ can be seen as the Hamiltonian of
a spin glass system [47]. Such a system can behave in a highly chaotic
way, hence giving rise to a large number of local minima. 
Indeed, such local minima, which correspond
to Gribov copies in the continuum gauge theory, appear
on the lattice, both for Abelian and non-Abelian fields, in the
Coulomb gauge as well as in the Landau gauge. The first numerical
evidence for lattice Gribov copies was presented in the early
nineties [41], [47], and [48]. The number of copies in the Coulomb
gauge was found to be larger than in the Landau gauge [48].

The effects on the evaluation of gauge-dependent
quantities resulting from the presence of Gribov copies are studied
by comparing the results for quantities, such as gluon and ghost
propagators [49], using two different averages: the average
considering only the absolute minima, which should give the result
in the minimal Landau gauge; and the average considering only the
first gauge-fixed copy generated for each configuration. If Gribov
copies were not considered, one would obtain 
the latter average as the result.

A very interesting analysis is provided by the evaluation of the
axial current renormalization constant $Z_{A}$ [41]. The relevance
is here twofold: on the one hand, $Z_{A}$ can be obtained from
chiral Ward identities in two distinct ways, either gauge
independent (it consists in evaluating matrix elements between
hadron states) or gauge-dependent (which consists in evaluating
matrix elements between quark states). This engenders an explicit
gauge-invariant estimate of $Z_{A}$, without any Gribov noise (in
that the Gribov copies consist of fluctuations giving rise to a
background noise), which can be directly contrasted with the
gauge-dependent estimate affected by Gribov copies. On the other
hand, there is also the advantage of finding $Z_{A}$ by solving an
algebraic equation of first degree for each time section of
the lattice, hence avoiding the systematic errors usually emerging
from an exponential fit of decay signals.

The corresponding analysis shows that Gribov copies have visible
effects, that can be detected, for example, on looking at the slightly
different estimates of $Z_{A}$ as a function of a parameter. The
Gribov fluctuations, however, i.e. those induced from the choice of
a particular Gribov copy, are small and do not prevail on the
statistical uncertainty. In other words, numerical effects of Gribov
copies on a lattice can be divided into two categories:
{\it measurement distortion} and {\it lattice Gribov noise}.

In most investigations of the influence of Gribov copies on lattice
quantities, it has been found that Gribov noise is of the same order as the
numerical accuracy of the simulations, and that it scales down as
a pure statistical error. In some particular cases Gribov noise
seems to be quite large. In maximally Abelian gauge it introduces
a clear bias on the number of monopoles [50] and on the value of
the Abelian string tension [51,52].

A typical example of distorsion resulting from Gribov copies is
the measurement of photon propagator in compact $U(1)$ lattice
gauge theories. It has been shown numerically [53--56], that
the photon propagator in the Coulomb phase is strongly affected by
Gribov noise and that only averages taken on absolute minima of
the minimizing functional reproduce the theoretical predictions.
Thus, one expects that the same might happen for the gluon and 
ghost propagators. Analogue results were found in pure $SU(2)$
lattice gauge theory in the minimal Landau gauge [43].

In some circumstances, in lattice too there exist copies 
which do not seem to be
related by a change in the winding index or from a singular local
gauge transformation; they are called non-topological Gribov
copies [54].

Gribov copies seem to increase in number with increasing lattice volume; 
experimental data, instead, suggest that some Gribov copies 
are lost as $\beta$ is
increased, at fixed volume. This is the behaviour that one 
would expect: a large lattice
volume, in fact, is related with a large number of gauge 
transformation variables
$D(x)$. Of course, when  $F_{L}[D]$
depends on more variables, it will have more local minima. 
Moreover, small $\beta$ means large coupling [46]; in the continuum limit we
may expect Gribov copies in such a situation. 
To obtain the continuum limit of
the theory we have to let both parameters go to infinity, hence
it appears highly plausible that the phenomenon of Gribov copies
described by lattice theories should occur in the continuum limit as
well [47].

It seems fair enough to say, however, that even for lattice theories
no undisputable results exist as yet. 
Investigation of the Gribov problem on a lattice has shown that
ambiguities resulting from the gauge-fixing procedure exist for
both Abelian and non-Abelian theories [57] in the Coulomb and
Landau gauges. For example, although on the
one hand there is evidence that Gribov copies exist in the $SU(3)$
Landau gauge [47], on the other hand the measurement of the $SU(3)$
gluon propagator obtained with the help of numerical simulations
performed in the Landau gauge does not point out the existence of
the Gribov ambiguity [43].

Anyway, every lattice calculation has to face the Gribov problem,
for example by employing stochastic gauge fixing, as is the case for
lattice evaluation of gluon screening masses [58].
Furthermore, on looking at the foundations of non-perturbative QCD,
if one takes locality and BRST symmetry as guiding principles [59], 
one again finds the unavoidability of Gribov copies, which are
instead absent at the price of dealing with non-local field theory [10].

\acknowledgments
The work of G. Esposito and F. Zaccaria has been partially
supported by PRIN 2002 {\it SINTESI}. We are grateful to
Gennadi Sardanashvily for encouragement, and to Mario Abud for
critical remarks.

\end{document}